\documentclass[12pt]{article}

\begin{document}

\begin{center}

{\baselineskip=24pt
{\Large {\bf The derivation of the dual superconductor theory from the
 Maximal Center projected $SU(3)$ -- gluodynamics} }

{\baselineskip=16pt
\vspace{1cm}

{ M.A.~Zubkov$^a$ }\\

\vspace{.5cm}
{ \it

$^a$ ITEP, B.Cheremushkinskaya 25, Moscow, 117259, Russia }} }
\end{center}
\begin{abstract}
We consider the Center projected $SU(3)$ gluodynamics and rewrite
 it as a dual superconductor theory. The center monopole field
 plays the role of Higgs field in the dual superconductor theory.
 The center monopole creation operator is constructed.
\end{abstract}
\newpage

\section{Introduction}

The investigation of the confinement problem is now one of the important
subjects of QCD. One of the most popular schemes of confinement is the
dual superconductor mechanism. According to this mechanism the quark-
antiquark string appears in the dual representation of the theory in the
way similar to the appearance of the Abrikosov string in Ginsburg - Landau
theory. But now it is not clear how the analogue of the Ginsburg - Landau
theory can be derived from the QCD.

The relativistic analogue of Ginsburg - Landau theory is the abelian Higgs
model. Thus the dual superconductor theory should be the nonlocal variant
of abelian Higgs model. There exists the point of view that the field
correspondent to the Maximal Abelian monopoles \cite{Wiese} should play
the role of Higgs field. The fact that the condensate of that
monopoles is the order parameter and is not equal to zero in the confinement
phase partially justifies this hypothesis. This picture seems to be quite
natural in the $SU(2)$ theory, but in the $SU(3)$ theory two maximal
abelian monopoles appear and the dual superconductor mechanism becomes too
complex.

Recently the approach alternative to the Maximal Abelian projection was
investigated. That approach is the Maximal Center projection. It occurs
that the Center dominance also takes place as the Abelian dominance
 \cite{Greensite}. And
 furthermore in the Maximal Center projected theory also one can construct
 the monopoles which are condensed in the confinement phase
 \cite{su3} \cite{su2}.
 The pleasant feature of that monopole is that there is only one center
monopole in the $SU(3)$ theory.

In this work we derive the representation of the Maximal Center projected
gluodynamics which has the form of the dual superconductor theory. The
field of the center monopole plays the role of the Higgs field. The quark
- antiquark string appears as the topological defect in this theory.
Also we construct the monopole creation and annihilation operators.

\section{The Maximal Center Projection.}
The Maximal Center projection is  the procedure of
 partial gauge fixing.
The gauge ambiquity is used to make all the link variables
$U \in SU(3)$ as close as possible
 to the elements of the center $Z_3$ of $SU(3)$:

$ Z_3 = \{{\rm diag}(\mathrm{e}^{(2\pi i /3) N}, \mathrm{e}^{(2\pi i /3) N},
 \mathrm{e}^{(2\pi i /3) N}\}$, where $N \in \{1, 0, -1\}$.

There is a lot of versions of the Maximal Center projection. That versions
differ from each other by the choice of the gauge fixing potential $O(U)$.
The projection is achieved by the minimizing of the potential $O(U)$
with respect to the gauge transformations $U_{xy} \rightarrow
g_x U_{xy} g_y^{-1}$.

This gauge condition is invariant under the central subgroup $Z_3$ of $SU(3)$.
Thus the Maximal center projected theory is nonlocal $Z_3$ gauge theory.
The well - known center vortices are defined as follows.
After fixing the Maximal Center gauge we define the
integer-valued link variable $N$:
\begin{eqnarray}
N_{xy}=0 &{\rm if}& ({\rm Arg}(U_{11})+{\rm Arg}(U_{22})+{\rm Arg}(U_{33}))/3
 \in \; ]-\pi/3, \pi/3], \nonumber \\
N_{xy}=1 &{\rm if}& ({\rm Arg}(U_{11})+{\rm Arg}(U_{22})+{\rm Arg}(U_{33}))/3
 \in \; ]\pi/3, \pi], \nonumber \\
N_{xy}=-1 &{\rm  if}& ({\rm Arg}(U_{11})+{\rm Arg}(U_{22})+{\rm Arg}(U_{33}))/3
 \in \; ]-\pi, -\pi/3].
\end{eqnarray}
In other words $N=0$ if $U$ is close to $1$,  $N=1$ if $U$ is close to
\noindent
$\mathrm{e}^{2\pi i/3}$ and  $N=-1$ if $U$ is close
 to $\mathrm{e}^{-2\pi i/3}$. Next we define the plaquette
variable:
 \begin{equation}
\sigma_{xywz} =  N_{xy}+N_{yw}-N_{zw}-N_{xz}
\end{equation}

In terms of the calculus of the differential forms on the lattice \cite{forms}
this equation looks like
\begin{equation}
\sigma = d N
\end{equation}

Then we introduce the dual lattice and define the variable $\sigma^*$
dual to $ \sigma$: if plaquette $^*\Omega$ is dual to plaquette $\Omega$, then
$\sigma^*_{^*\Omega} = \sigma_{ \Omega}$.  One can
easily check that the variable $\sigma$ represents a closed
surface. This surface  is known as the worldsheet of the center
vortex.

We express the $SU(3)$
gauge field $U$  as the product of $\exp((2\pi i / 3) N)$  and $V$,
where  $V$ is the $SU(3)/Z_3$ variable
$({\rm Arg}(V_{11})+{\rm Arg}(V_{22})+{\rm Arg}(V_{33}))/3 \in \; ]-\pi/3,
\pi/3]$. Then
$U=\exp((2\pi i/ 3) N) V$

After that we represent the action of the Wilson loop $C$ as follows:
\begin{equation}
 W_C = \Pi_C U =\exp((2\pi i / 3) L(C,\sigma)) \Pi_C V
\end{equation}
The term $(2\pi i/3)
L(C,\sigma)$ is known as the Aharonov - Bohm interaction term.
The quantity $L(C,\sigma)$ is the linking number of the loop $C$ and the
closed surface $\sigma^*$.

The  center dominance means  that after the
Maximal Center projection the Aharonov - Bohm interaction term
causes confinement and produces the full string tension.

The center monopole is just the $Z_3$ analogue of the monopole in $U(1)$
theory. Let us recall that monopoles in $U(1)$ theory
are constructed as loops on which the force lines of the
gauge field end. It is well known that in electrodynamics the Maxwell
equations $dF=0$ restrict the existence of magnetic charges. But
in the compact theory values of F which differ from each other
by $2\pi$, are equivalent. Thus the correct field strength is
$F \, {\rm mod} \, 2\pi$ and $^*d (F \, {\rm mod} \, 2\pi) = 2\pi j_m$,
where $j_m$ is the monopole current.

In the $Z_3$ theory $\sigma=dN$ is the analogue of the field strength
$F=dA$. The Aharonov - Bohm interaction between the center vortex and the
quark depends
only on $[\sigma] \, {\rm mod} \, 3$.
The variable $[\sigma] \, {\rm mod} \, 3$ represents the surface with boundary.
This boundary is a closed line. We assume that this line represents
the world trajectory of the particle, which we call a center monopole:
\begin{equation}
 3j_m = ^*d ([\sigma] \, {\rm mod} \, 3) =
 \delta  ([\sigma^*] \, {\rm mod} \, 3).
\end{equation}

\section{The derivation of the dual superconductor theory}

We start with the Wilson loop average in the $SU(3)$ theory.

\begin{equation}
< W(C) > = \int D U exp(-\sum_{plaq} \beta (1-1/3 Re Tr U_{plaq})Re Tr \Pi_C U
\end{equation}

To consider the Maximal Center projection we have to use
the Faddeev-Popov unity

\begin{equation}
1=\lim_{\alpha \rightarrow \infty} \int Dg exp(-\alpha O(g_x U_{xy}
g^{-1}_y))\Delta_{FP}(U,\alpha)
\end{equation}

Here the gauge condition is that we assume $U$ to provide a minimum
of the functional $O(U)$.

This functional is the given measure of the distance in the functional
 space between  the matrix function $U$ and the set of functions,
 which attach one
 of the center elements of $SU(3)$ to each link.
It is obvious, that the functional $O$ is invariant
 under the remaining $Z_3$ symmetry.

As in the previous section we consider the following representation
of the projected $U$;

\begin{equation}
U = e^{2\pi i N/3} V
\end{equation}

where $V$ is closer to $1$ than to other center elements.

It follows from the invariance of $O$ under the $Z_3$ symmetry, that the
Faddeev-Popov
determinant does not depend upon $N$.

Thus we have
\begin{eqnarray}
< W(C) > = lim_{\alpha \rightarrow \infty }\int D_{V\in SU(3)/Z_3} V
\sum_{N=1,0,-1}
\nonumber\\ exp(-\sum_{plaq} \beta (1-1/3 Re Tr e^{2\pi dN/3} V_{plaq})
\nonumber\\-\alpha O(V)+(2\pi i/3) (N,C))
 \Delta_{FP}(V,\alpha) Re Tr \Pi_C V
\end{eqnarray}

The center dominance means, that we can omit the expression
$Re Tr \Pi_C V$ to calculate the
correct string tension.
Thus we are considering the following expression for the $Z_3$ Wilson loop:

\begin{eqnarray}
< Z(C) > = lim_{\alpha \rightarrow \infty} \int D_{V\in SU(3)/Z_3} V
\sum_{N=1,0,-1}
\nonumber\\ exp(-\sum_{plaq} \beta (1-1/3 Re Tr e^{2\pi dN/3} V_{plaq})
\nonumber\\-\alpha O(V)+(2\pi i/3) (N,C))
 \Delta_{FP}(V,\alpha)
\end{eqnarray}
After the integration over short - ranged field $V$ we obtain the
resulting $Z_3$ theory with nonlocal action
\begin{equation}
< Z(C) > =\sum_{N=1,0,-1} exp(-  S(dN mod 3)+(2\pi i/3) (N,C))
\end{equation}
It seems that this theory behaves like an usial $Z_3$ theory in the
confinement phase.

Now we are going to transfer ourselves into the dual representation
 of the above theory
to understand how  the superconductor appears.
First let us remember, that the center monopoles are defined as
$j = 1/3  ^*d(dN mod3)$.

Then we apply the dual transformation.

\begin{eqnarray}
< Z(C) > =\sum_{N=1,0,-1}\sum_{m=1,0,-1}\sum_{j,n,\delta j = 0} exp(-  S(dN
mod 3)\nonumber\\+(2\pi i/3) (N,C))  \delta(m-dN-3n)\delta(3^*j-d m)
\end{eqnarray}

We use the formulas
\begin{eqnarray}
\sum_{k=0,1,-1} e^{(2\pi i/3) (Z,k)} = \sum_n \delta(Z-3n);
\nonumber \\
\int_{-\pi}^{\pi} dh e^{i h Z} = \delta(Z),
\end{eqnarray}
to obtain:

\begin{eqnarray}
< Z(C) > =\sum_{N,m,k=1,0,-1}\sum_{j, \delta j = 0}\int_{-\pi}^{\pi} D h
exp(-  S(dN mod 3) \nonumber\\+(2\pi i/3) (N,C)+(m-dN,k)2\pi i/3 +
i(h,3^*j-dm))
\end{eqnarray}
Then we use the expression $C = \delta A[C]$, where $A$ is some
surface,
spanned on the quark loop.

\begin{eqnarray}
< Z(C) > =\sum_{N,m,k=1,0,-1}\sum{\delta j=0}\int_{-\pi}^{\pi} D h exp(-  S(m)
\nonumber\\+ (m, -\delta h +(2\pi /3) (A[C]+k))-(2\pi i/3)(N,\delta k) +
i(h,3^*j))
\end{eqnarray}

We can perform the summation over $N$ to obtain the constraint
 $\delta k = 3 l$ for some integer $l$.
 Also we can perform the summation over $m$, obtaining
\begin{equation}
exp(-Q(f)) = \sum_{m=1,0,-1} exp(-s(m) + i(m,f))
\end{equation}

It's obvious, that $Q(f)$ is periodic with the period $2\pi$.
Thus we get

\begin{eqnarray}
< Z(C) > =\sum_{k=1,0,-1; \delta k =3l;l;j}\int_{-\pi}^{\pi} D h exp(-
Q( -\delta h +(2\pi /3) (A[C]+k)))
\nonumber \\
exp( i(h,3j))
\end{eqnarray}
We can solve the constraint $\delta k = 3 l$ : $k = 3A[l]+\delta z $. Due
to the periodicity
of $Q$, $l$ is eliminated. Then we redefine $h \rightarrow (h+ 2\pi/3
\delta z)mod 2\pi$, and finally get:

\begin{equation}
< Z(C) > =\sum_{\delta j = 0}\int_{-\pi}^{\pi} D h exp(-  Q( -\delta h
+(2\pi /3) A[C]) + i(h,3^*j)) 
\end{equation}

Thus we have obtained that the theory dual to the original $Z_3$ projected
 gluodynamics
is just the nonlocal $U(1)$ gauge theory  with additional summation over
 the worldlines of the center monopoles, which carry the charge $3$ with
 respect to the mentioned $U(1)$ gauge field.

We can rewrite the summation over the worldlines of the monopoles as the
integral over
the Higgs field of charge $3$:

\begin{equation}
\sum_{j} exp(-i(H,3j))
 = \int D_{\Phi\in C}\Phi exp( -\sum_{xy} \Phi_{x} e^{3iH_{xy}} \Phi_y^+
- V(|\Phi|)),
\end{equation}
where the potential $V$ is infinitely deep, and the vacuum average of $|\Phi|$
is infinitely large. So $V(r)= a (r^2-b)^2$, where $a,b \rightarrow \infty $.
Here we denoted $ ^*h = -H$.

Finally we have
\begin{eqnarray}
 < Z(C) > =\int_{-\pi}^{\pi} D H\int D_{\Phi\in C}\Phi exp(
-Q(dH+(2\pi/3)^*A[C])\nonumber\\
 -\sum_{xy} \Phi_{x} e^{3iH_{xy}} \Phi_y^+ - V(|\Phi|))
\end{eqnarray}

The last representation is  the dual superconductor representation of the
$SU(3)$ theory.
Here the field $\Phi$ is the field of our center monopole.
When the monopole is condensed, the Abrikosov-Nielsen Olesen strings appear.
That strings carry the magnetic flow $2\pi/3$, thus connecting the quarks,
which play the role
of monopoles here.  Also the usial monopoles, existing due to the periodicity
 of the
action, create the Abrikosov-Nielsen-Olesen strings. But that strings should
carry
the magnetic flow $2\pi$, which is clear from the above expression.
Thus that monopoles indeed create $3$ strings and do not influence the
confinement
mechanism.

\section{The center monopole creation operator}

	We define the monopole creation operator in the way similar to that
of in the $U(1)$ theory representing the Maximal Abelian projected
$SU(2)$ - gluodynamics. Following  \cite{Polikarp}, we obtain the vacuum
average of the monopole - antimonopole correlator in the dual theory:

\begin{eqnarray}
< \Phi(z1)\Phi^*(z2) > =\int_{-\pi}^{\pi} D H\int D_{\Phi\in C}\Phi exp(
-Q(dH)\nonumber\\
 -\sum_{xy} \Phi_{x} e^{3iH_{xy}} \Phi_y^+
 - V(|\Phi|))\Phi(z1)\Phi^*(z2)exp(i (D(z1)-D(z2),H)),
\end{eqnarray}
where $ \delta D(z) = -3\delta_z$.

After coming back to the representation throw the worldlines
 of the monopoles we get

\begin{eqnarray}
< \Phi(z1)\Phi^*(z2) > =\sum_{\delta j=
\delta_{z1}-\delta_{z2}}\int_{-\pi}^{\pi} D H
 exp(-  Q( d H ) + \nonumber \\
i(H,3j-(D(z1)-D(z2)))
\end{eqnarray}

Repeating the steps back to  the original representation we obtain

\begin{eqnarray}
<\Phi(z1)\Phi^*(z2)> = lim_{\alpha \rightarrow \infty}
 \int D_{V\in SU(3)/Z_3} V \sum_{N=1,0,-1}
\nonumber \\
exp(-\sum_{plaq}
\beta (1-1/3 Re Tr
 e^{2\pi (dN)/3} V_{plaq})
\nonumber\\-\alpha O(V))
 \Delta_{FP}(V,\alpha)\nonumber \\
Q(^*d([dN]mod3)-(3j_{z1,z2}-D(z1)+D(z2)))
\end{eqnarray}

Here $j_{z1,z2}$ is the line connecting points
 $z1$ and $z2$ of the dual lattice.
And

\begin{equation}
Q(x) = \sum_{\delta j = 0} \Pi_{^*links}sin(\pi(x-j))/(x-j) ,
\end{equation}

where the summation is over the closed integer 1-forms on the dual lattice.

In other words
\begin{eqnarray}
<\Phi(z1)\Phi^*(z2)> =
 \int D U  exp(-\sum_{plaq}
\beta (1-1/3 Re Tr U_{plaq}))
\nonumber\\
 Q(3^*j_{Z_3}-(3j_{z1,z2}-D(z1)+D(z2))),
\end{eqnarray}
where $j_{Z_3}$ is the center  monopole trajectory
extracted from the field configuration $U$.

\section{Conclusions}

	In this work we made an attempt to extract the kind of superconductor
theory from the $SU(3)$ gluodynamics in the Maximal Center Gauge. We
obtained the theory, which contains $U(1)$ gauge field and the scalar
 field charged with respect to that $U(1)$ field. The action is essentially
 nonlocal. The usial quarks play the role of monopoles in this theory. The
scalar field is condensed in the confinement phase and is not condensed in
the deconfinement phase. The worldlines of the particle correspondent to that
 scalar field are just the worldlines of the center monopoles.

Of course, the analogous representation one can obtain for the
gluodynamics in the Maximal Abelian Gauge. But in that case two
 scalar fields and two $U(1)$ fields appear. Thus the dual superconductor
mechanism becomes too complex and unnatural.

\section*{Acknowledgments}

	The author is grateful to  B.L.G. Bakker, J. Greensite,
 and  A. Veselov for useful discussions.

This work was supported by the JSPS Program on Japan--FSU scientists
collaboration, by the grants
INTAS-RFBR-95-0681 and RFBR-97-02-17491.

\end{document}